\documentclass[conference]{IEEEtran}
\usepackage{latexsym}
\usepackage{amsmath}
\usepackage{amssymb}
\usepackage{amsthm}

\usepackage{graphicx}
\usepackage{epsfig}
\usepackage{amssymb}
\usepackage{subcaption}
\usepackage{mwe}

\usepackage{float}
\usepackage{mathtools}
\usepackage{algorithm}
\usepackage[noend]{algpseudocode}
\usepackage[utf8]{inputenc}
\usepackage[english]{babel}
\usepackage{fancyhdr}
\usepackage{mleftright}
\usepackage{courier}
\usepackage{array}
\usepackage{booktabs}
\usepackage{syntonly}
\usepackage{xcolor}
\usepackage{colortbl}
\usepackage{lipsum}
\usepackage{epstopdf}
\usepackage{environ}
\usepackage{relsize}
\usepackage{multirow}
\usepackage{diagbox}
\usepackage{color} 
\usepackage[normalem]{ulem} 
\usepackage[makeroom]{cancel} 
\usepackage{bm}
\usepackage{enumitem}
\usepackage[normalem]{ulem}

\NewEnviron{myequation}[1][]{%
\begin{equation}
\scalebox{#1}{$\begin{aligned}\BODY\end{aligned}$}
\end{equation}
}
\NewEnviron{mymultiline}[1][]{%
\scalebox{#1}{\begin{align}
\BODY
\end{align}}
}
\newcommand{\vb}{\boldsymbol}

\newcommand{\m}[1]{\mathrm{#1}}
\newcommand{\mc}[1]{{\mathcal{#1}}}

\newcommand{\specialcell}[2][c]{%
  \begin{tabular}[#1]{@{}c@{}}#2\end{tabular}}

\newcommand{\cmmnt}[1]{\ignorespaces}

\newcommand{\PreserveBackslash}[1]{\let\temp=\\#1\let\\=\temp}
\newcolumntype{C}[1]{>{\PreserveBackslash\centering}p{#1}}
\newcolumntype{R}[1]{>{\PreserveBackslash\raggedleft}p{#1}}
\newcolumntype{L}[1]{>{\PreserveBackslash\raggedright}p{#1}}

\makeatletter
\let\OldStatex\Statex
\renewcommand{\Statex}[1][3]{%
  \setlength\@tempdima{\algorithmicindent}%
  \OldStatex\hskip\dimexpr#1\@tempdima\relax}
\makeatother

\newtheorem{theorem}{Theorem}
\newtheorem{lemma}{Lemma}

\ifCLASSINFOpdf

\else

\fi

\begin{document}
\setlength{\belowdisplayskip}{0.4\baselineskip} \setlength{\belowdisplayshortskip}{0.1\baselineskip}
\setlength{\abovedisplayskip}{0.4\baselineskip} \setlength{\abovedisplayshortskip}{0.1\baselineskip}

\title{Batch-Constrained Reinforcement Learning for Dynamic Distribution Network Reconfiguration}

\author{\IEEEauthorblockN{Yuanqi Gao, \textit{Student Member, IEEE}, Wei Wang, \textit{Student Member, IEEE}, Jie Shi, \textit{Student Member, IEEE}, \\ and Nanpeng Yu, \textit{Senior Member, IEEE}}}

\maketitle
\begin{abstract}



Dynamic distribution network reconfiguration (DNR) algorithms perform hourly status changes of remotely controllable switches to improve distribution system performance. The problem is typically solved by physical model-based control algorithms, which not only rely on accurate network parameters but also lack scalability. To address these limitations, this paper develops a data-driven batch-constrained reinforcement learning (RL) algorithm for the dynamic DNR problem. The proposed RL algorithm learns the network reconfiguration control policy from a finite historical operational dataset without interacting with the distribution network. The numerical study results on three distribution networks show that the proposed algorithm not only outperforms state-of-the-art RL algorithms but also improves the behavior control policy, which generated the historical operational data. The proposed algorithm is also very scalable and can find a desirable network reconfiguration solution in real-time.

\end{abstract}

\IEEEpeerreviewmaketitle

\section{Introduction}
With increasing penetration of remotely controllable switches and distributed generations (DGs), distribution network reconfiguration (DNR) \cite{Jabr2012reconfiguration} became critical in increasing the hosting capacity of distributed energy resources (DERs) \cite{Capitanescu2015Assessing}, minimizing the curtailment of DGs \cite{LeiDDNR2018}, and reducing network line losses \cite{Dorostkar2016Value}. DNR works by changing the status of switching devices \cite{Bilibin2014Comprehensive} to optimize certain operational objectives while satisfying operational constraints, which include the voltage magnitude limit and network radiality. DNR can be performed either statically or dynamically \cite{Capitanescu2015Assessing} \cite{Gao1910:Dynamic}. The former is concerned with determining the best static network configuration for the entire study period. The latter aims to find a sequence of network configurations over time. We focus on the dynamic DNR in this paper.

The DNR problem is typically formulated as a mixed-integer programming (MIP) problem, where the integer variables represent the status of remotely controllable switches. The problem size for dynamics DNR is typically much larger than the static ones. 
For the switch statuses of multiple time steps need to be identified in the dynamic DNR problems. Furthermore, it is more difficult to handle uncertainties associated with loads and DGs in dynamic DNR problems.

The existing literature on dynamic DNR can be categorized into three groups: the mixed-integer programming based approaches, the heuristic or meta-heuristic algorithms, and dynamic programming methods.

The first group of literature utilizes mixed-integer programming framework to formulate the dynamic DNR as a deterministic, stochastic, or robust optimization problem. Deterministic optimization formulations do not take stochastic power injections into consideration. The optimization methods used to solve deterministic problems include mixed-integer linear programming (MILP) \cite{novoselnik2015dynamic} \cite{Ahmed2019acdc}, mixed-integer conic programming (MICP) \cite{Dorostkar2016Value}, mixed-integer nonlinear programming (MINLP) \cite{Kianmehr2019resilience}, and MIP combined with other problem size reduction heuristics \cite{Capitanescu2015Assessing}. Unlike deterministic optimization, stochastic and robust optimization methods fully incorporate the uncertainties of loads and DGs into the problem formulation. Robust optimization methods are developed to find the reconfiguration with optimal performance in the worst-case scenario \cite{Lee2015Robust} \cite{Haghighat2016Robust} \cite{Akrami2020muPMU} and simultaneously identify the critical switch \cite{LeiDDNR2018}. Stochastic optimization methods are developed to optimize the expected control objective \cite{Dantas2017dynamic}, or incorporate the uncertainties of the loads and DGs by combing MILP with unscented transforms \cite{Fard2018effective}.

The second group of literature uses heuristics or meta-heuristic algorithms. The minimum spanning tree \cite{mosbah2017optimum} and the branch exchange \cite{bernardon2014real} methods are used to heuristically solve the dynamic DNR problem. Meta-heuristic algorithms such as genetic algorithm \cite{Liu2019intraday}, 
fuzzy adaptive inference-based particle swarm optimization (PSO) \cite{Chen2016Comprehensive}, and a hybrid PSO with time-partitioning \cite{Fu2018toward} have been adopted to identify the optimal network configurations.

The third group of literature leverages dynamic programming (DP) methods \cite{wu2003} to determine the optimal sequence of network configurations. This approach first identifies the set of radial configurations and treats them as the states. It then applies the DP backward iteration \cite{feinberg2011rolling} to determine the optimal sequence of hourly network configurations.

Most of the existing literature uses a physical model-based control approach to solve the dynamic DNR problem. However, this approach has two limitations. First, model-based algorithms may not be reliable when electric utilities do not have complete and accurate distribution network parameters. It is well known that it is difficult for electric utilities to maintain accurate  primary and secondary feeders' parameters for distribution networks covering millions of nodes \cmmnt{\cite{wang2016phase}, \cite{foggo2018comprehensive},} \cite{YGao2019Physically}. Second, the computation time for model-based control algorithms increases exponentially with the number of remotely controllable switches, the number of DERs, and the length of the operation horizon, which makes it difficult to apply in real-time network reconfiguration.

To address these limitations, we use a data-driven approach to formulate the dynamic DNR as a reinforcement learning (RL) problem. In a typical RL setup, an agent tries to learn an optimal control policy by interacting with the real physical environment or a simulated one. However, it is costly and time consuming for the agent to learn an optimal network reconfiguration strategy by directly interacting with the physical distribution network. Furthermore, it is difficult to create a reliable simulated environment when the network parameters are inaccurate. Thus, it is desirable for the agent to learn from the historical network reconfiguration data collected by the electric utilities. Nonetheless, learning a control policy from a finite dataset brings about its own challenge. The network reconfiguration problem has an extremely large number of feasible solutions and reconfiguration actions. Most of them are never visited in the historical dataset. Directly applying state-of-the-art RL algorithms can induce large extrapolation errors \cite{fujimoto2019without}. Consequently, the trained control policy from the RL algorithms can under-perform the behavior control policy that generated the data.

In this paper, we develop a novel RL algorithm called batch-constrained soft actor-critic (BCSAC) to overcome the extrapolation error problem. Our proposed algorithm trains a control policy, which maximizes the total discounted return while minimizing the dissimilarity between the learned control policy and the behavior policy of the batch data. Our BCSAC algorithm represents the behavior policy with a conditional variational autoencoder and regularizes the reward function with the Kullback–Leibler (KL) divergence between the learned control policy and the behavior policy. We prove the convergence of the KL-divergence regularized (batch-constrained) version of the policy iteration. Then we train the BCSAC algorithm using standard machine learning optimization routine. The trained algorithm is then
evaluated on several test distribution networks with real-world smart meter data. Numerical study results show that our proposed BCSAC algorithm is able to successfully learn a network reconfiguration strategy without interacting with the distribution network. It not only improves the behavior control policy but also outperforms state-of-the-art off-policy RL algorithms.

The unique contributions of this paper are listed below:

$\bullet$ This paper proposes a novel BCSAC algorithm to learn effective control policies from a finite historical operational dataset.

$\bullet$ This paper provides the convergence proof of the off-policy batch-constrained policy iteration.

$\bullet$ This is the first data-driven dynamic DNR algorithm that improves a baseline control policy for test distribution networks with billions of feasible configurations, without network parameter information and direct interaction with the real-world distribution system.

$\bullet$ The proposed BCSAC algorithm not only yields much lower network operational costs than state-of-the-art off-policy RL algorithms but also has a much shorter computation time than model-based control algorithms.

The rest of the paper is organized as follows: Section II formulates the dynamic DNR problem as a Markov decision process. Section III presents the technical methods of our proposed BCSAC algorithm. Section IV shows the numerical study results. Section V states the conclusion.

\section{Problem Formulation}\label{section2}
In this section, we present our formulation for the dynamic DNR problem as a Markov decision process (MDP) \cite{sutton2018reinforcement}. First we review the preliminaries of MDPs. Next we describe the dynamic DNR problem as an MDP. Finally we state the set up of the reinforcement learning problem for DNR.

\subsection{Basics of Markov Decision Process}\label{section2A}
An MDP $\mc{M} = (\mc{S}, \mc{A}, p, r, \gamma, T)$ consists of a state space $\mc{S}$, an action space $\mc{A}$, a state transition probability $P(s^\prime|s,a)$ $\forall s^\prime,s\in\mc{S}, \forall a\in\mc{A}$, a reward function $r(s,a): \mc{S}\times\mc{A}\mapsto \mathbb{R}$, $\forall s\in\mc{S}$, $\forall a\in\mc{A}$, a discount factor $\gamma\in[0,1)$, and a time horizon $T$. In an MDP, an agent selects an action $A_t\in \mc{A}$ based on the environment's state $S_t\in\mc{S}$ at each discrete time step $t$. Then the agent receives a reward $R_{t+1} = r(S_t, A_t)$ and the environment's state transitions to $S_{t+1}$ according to the state transition probability $P(S_{t+1}|S_t, A_t)$. The process either terminates when $t=|T|$ if $T$ is finite or continues indefinitely if $T$ is infinite. 

The goal of the agent is to find a control policy $\pi$ that maximizes the expected discounted return 
$J(\pi) = \mathbb{E}_{\tau \sim \pi}[G(\tau)]$, where control policy $\pi(\cdot|s)$ maps each state to an action selection probability distribution over the action space $\mc{A}$. $\tau$ is a trajectory or sequence of states and actions, $\{S_0,A_0,S_1,A_1,...,S_{T-1},A_{T-1},S_T\}$. $G(\tau)$ is the discounted return along a trajectory. $G(\tau) = \sum_{t=0}^T \gamma^t R_{t+1}$.

Finally, we define two important value functions, the state-value function $v_\pi(s)$ and the action-value function $q_\pi(s,a)$ with respect to the control policy $\pi$:
\begin{align}
v_\pi(s) &= \underset{\tau \sim \pi}{\mathbb{E}}\left[\textstyle \sum_{k=0}^T\gamma^kR_{t+k+1}|S_t=s\right] \quad \forall s \label{eq1}\\
\hspace*{-1cm}q_\pi(s,a) &= \underset{\tau \sim \pi}{\mathbb{E}}\left[\textstyle \sum_{k=0}^T\gamma^kR_{t+k+1}|S_t=s,A_t=a\right] \forall s,a \label{eq2}
\end{align}
where $v_\pi(s)$ and $q_\pi(s,a)$ represent the expected discounted return starting from state $s$ or state-action pair $(s,a)$, and following control policy $\pi$ thereafter. Next we formulate the dynamic DNR problem as an MDP.

\subsection{Formulate Dynamic DNR as an MDP}
We consider a distribution network with $n$ load nodes, $m$ lines, and $n_S$ substations. Let $v_{it}$, $p_{it}$, and $q_{it}$ denote the voltage magnitude, real and reactive power net injections of node $i$ at time $t$. $p^l_t$ denotes the network's total real line losses. The binary variable $\alpha_{\ell t}$ represents the status of the switch $\ell$. $\alpha_{\ell t}=1$ if switch $\ell$ is closed at time $t$. We define vectors for nodal real and reactive power injections and branch status at time $t$ as $\vb{p}_t=[p_{1t}, \cdots, p_{nt}]$, $\vb{q}_t=[q_{1t}, \cdots, q_{nt}]$, and $\vb{\alpha}_t=[\alpha_{1t},\cdots,\alpha_{mt}]$.

Now, we formulate the dynamic DNR problem as an MDP by identifying the agent, state, action, and reward. The agent is the distribution system operator or controller. The state at time $t$ is defined as $S_t = [\vb{p}_t,\vb{q}_t,\vb{\alpha}_t,t]$. Thus, $\mc{S}$ consists of the set of all power injection patterns together with the set of all radial configurations. We define the action taken at time $t$, $A_t$, as changing the topology of the network by a single pair of branch status exchange, that is, closing a switch in $\{1,\cdots,m\}$ and opening another one, such that the resulting configuration $\vb{\alpha}_{t+1}$ is still radial \cite{civanlar1988distribution}\cmmnt{\cite{baran1989network}, \cite{khodabakhsh2017submodular}}. We deem opening and closing the same switch as staying in the same configuration, and it does not incur a switching cost. Note that starting from a given configuration $\vb{\alpha}_{t}$, only a subset of all switch pairs is feasible; the others will result in a loop or disconnected network. Thus, in each state $s$, only a subset of actions are allowed to be chosen. We defer the implementation details to Section III.H.

The above formulated states and actions uniquely define a state-transition probability model $P(S_{t+1}|S_t, A_t)$ between the current state $S_t = [\vb{p}_t, \vb{q}_t, \bm{\alpha}_t, t]$ and the next state $S_{t+1}$. The new state's variables $[\vb{p}_{t+1}, \vb{q}_{t+1}, \bm{\alpha}_{t+1}, t+1]$ are determined by $S_t$ and $A_t$ as follows. The transition probability between adjacent time steps' power injections $P(\vb{p}_{t+1},\vb{q}_{t+1}|\vb{p}_t, \vb{q}_t)$ can be described by the random process of the power injections and is not affected by the action. $\bm{\alpha}_{t+1}$ is determined by $\bm{\alpha}_{t}$ and the open/closing switches in $A_t$. The global time variable $t$ is increased by $1$.

The reward function reflects the cost associated with network line losses $p^l_t$, the switching cost, and operating limit constraint violation penalty and is defined as follows.
\begin{align}\label{eq3}
R_{t+1} = r(S_t,&A_t) = -C^lp^l_t(\vb{p}_t,\vb{q}_t,\vb{\alpha}_{t+1}) \nonumber \\ 
 & - C^s|\vb{\alpha}_{t+1}-\vb{\alpha}_t| - \lambda c(\vb{p}_t,\vb{q}_t,\vb{\alpha}_{t+1})
\end{align}
where $C^l$ is the unit cost of electricity. $C^s$ is the cost of opening or closing of a switch. The third term in (\ref{eq3}) describes the voltage constraint violation penalty \cite{Gao1910:Dynamic}:
\begin{align}
c(\vb{p}_t,\vb{q}_t,\vb{\alpha}_{t+1}) = \sum_{i\in N^v}[\m{max}(0,v_{it}-\bar{v}) +\m{max}(0,\underline{v}-v_{it})]
\end{align}
where $N^v$ is the set of all nodes that have voltage measurement devices; $\bar{v}$ and $\underline{v}$ are the upper and lower bounds for voltage; $\lambda$ is the penalty factor associated with the voltage constraint violation. The value of $\lambda$ can be determined based on operational considerations and empirical performance. The exact value of $\lambda$ will be provided at Section IV.A.

Finally, we choose a discount factor $\gamma$ that is less than 1  and set $T=\infty$. This completes the MDP formulation for the dynamic DNR problem. In sum, the dynamic DNR problem is a continuing task with a finite action space and a state space with continuous variables. The learning setup of this MDP is explained in the next subsection.

\subsection{The Learning Setup}
In general, there are two types of RL methods to solve MDPs, on-policy RL and off-policy RL. If we develop an on-policy RL method to solve the dynamic DNR problem, then we must evaluate and improve the control policy that is used to make network reconfiguration decisions. However, this is inappropriate for both the real-world and the simulation environment. It is not only costly but also risky to apply an insufficiently trained control policy on real-world distribution network \cite{garcia2015comprehensive}. Furthermore, most regional electric utilities do not have accurate and reliable network parameter information \cmmnt{\cite{wang2016phase}, }\cite{YGao2019Physically}. Thus, evaluating and improving the control policy through interaction with an inaccurate simulation model for distribution network is not desirable.

In this paper, we take the off-policy RL approach to improve a control policy different from that used to generate the historical operational data. The RL task for the dynamic DNR problem is to learn a good control policy from a given set of historical operational data. The setup of learning from a given historical dataset rather than from directly interacting with the environment is known as batch reinforcement learning \cite{lange2012batch}. From now on, the term batch and historical operational data will be used interchangeably. The historical operational data should contain relevant information about the state, action, and reward of the MDP and will be explained in detail below.

We assume the historical operational data are to be derived through the following measurements collected by an electric utility. First, the nodal power injections $p_{it}+jq_{it}$ at each time step and node with non-zero injection are recorded by smart meters or other sensors. Second, the SCADA system records real and reactive power at the substations. Third, the nodal voltage magnitude data $v_{it}$ is available from the SCADA system at a subset of nodes in the network. Finally, the switch status $\vb{\alpha}_t$ are available from the remotely controllable switches. With these measurements, we can construct the historical states, actions, and rewards of the MDP. In particular, the network loss can be estimated as the sum of all net power injections of the distribution network $p^l_t = \sum_{i=1}^{n+n_S} p_{it}$.

Two factors make it challenging to develop a batch RL algorithm to solve the MDP representing the dynamic DNR problem. First, the state space of the MDP is high-dimensional and grows exponentially with the size of the distribution network. Leveraging function approximators such as neural networks to estimate the value function or control policy associated with this high-dimensional state space is not straightforward. Second, the batch RL controller can only learn from the limited information contained in a finite amount of historical operational data.

\section{Technical Methods}
In this section, we first present the preliminaries of RL algorithms and batch RL algorithms. Then we develop our proposed BCSAC algorithm. Finally, we provide the RL algorithm implementation details for the dynamic DNR problem.

\subsection{Basics of Reinforcement Learning}
The off-policy RL algorithms can be categorized into two groups, action-value methods and policy gradient methods. Action-value methods such as deep Q network (DQN) approximate the action-value functions through learning and then select actions based on the estimated action-value functions. However, for RL problems such as dynamic DNR with extremely large state-space and action-space, it can be difficult to approximate the action-value functions. To deal with these problems, researchers developed the policy gradient methods, which learn a parameterized control policy that directly selects actions without consulting a value function \cite{sutton2018reinforcement}. To further improve the sample efficiency and robustness of the policy gradient methods, state-of-the-art maximum entropy RL algorithms such as soft actor-critic (SAC) \cite{haarnoja2018soft} have been developed. Next, we provide a brief review of the SAC algorithm.

\subsubsection{Soft Actor-Critic}
Soft actor-critic \cite{haarnoja2018soft} regularizes the reward function by the entropy of the policy: $r(s,a)+\tau H(\pi(\cdot|s))$, whose contribution to the reward is controlled by the temperature parameter $\tau$. The entropy regularized state-value functions $v_\pi^{\m{h}}(s)$ and action-value functions $q_\pi^{\m{h}}(s,a)$ are shown to satisfy \cite{Nachum2017Bridging}:
\begin{align}\label{eq11}
v_\pi^{\m{h}}(s) = \mathbb{E}_{a\sim\pi}\mathbb{E}_{s^\prime\sim P}\left[ r+\gamma v_\pi^{\m{h}}(s^\prime)\right] + \tau H(\pi(\cdot|s))
\end{align}
\begin{align}\label{eq12}
q_\pi^{\m{h}}(s,a) = r + \gamma\mathbb{E}_{s^\prime\sim P}\left[ v_\pi^{\m{h}}(s^\prime)\right]
\end{align}
\begin{align}\label{eq13}
v_\pi^{\m{h}}(s) = \mathbb{E}_{a\sim\pi}[q_\pi^{\m{h}}(s,a)] + \tau H(\pi(\cdot|s))
\end{align}

To deal with large continuous domains, the value functions (critic) and the policy function (actor) shown above can be approximated by neural networks: $v_\psi(s), q_\theta(s,a), \pi_\phi(a|s)$, where $\psi$, $\theta$, and $\phi$ are the parameters of the corresponding neural networks.

The SAC algorithm works by iteratively updating the parameters of the value functions and the policy function. The parameters of value functions can be updated according to the gradient of the squared residual error of state-value function and the soft Bellman residual of action-value function. The parameters of the policy can be updated by
\begin{align}\label{eq14}
	\hspace*{-0.35cm} \pi_{new}(\cdot|s)=\arg\min_{\pi}\; D_\m{KL}\left( \pi(\cdot|s)||\frac{\m{exp}(q^\m{h}_{\pi_{old}}(s,a)/\tau)}{Z_{\pi_{old}}(s)}\right)
\end{align}
where $Z_{\pi_{old}}(s)$ is the partition function that normalizes the numerator to a probability distribution. $D_{\m{KL}}(p||q)$ is the KL-divergence between distributions $p$ and $q$.

\subsection{Batch-constrained Reinforcement Learning}
In the batch RL setup, the agent can only learn from a finite dataset collected by some sampling procedure. For example, the historical operational dataset may be generated from a model-based controller and/or heuristic control actions selected by operators. Therefore, if we directly apply off-policy RL algorithms such as DQN or SAC in the batch RL setup, then the action-value function $q_\pi(s,a)$ of a given policy $\pi$ may not be accurately evaluated. As a result, the learning agent may erroneously extrapolate $q_\pi(s,a)$ of some actions $a$ to higher values \cite{fujimoto2019without}. Formally, let $q_\pi(s,a)$ denote the true action-value function of a policy $\pi$ and $q_\pi^{\mc{D}}(s,a)$ denote the action-value function of policy $\pi$ estimated using the batch data. Then the extrapolation error of a state-action pair $\epsilon_\pi(s,a)$ and the extrapolation error of policy $\epsilon_\pi$ can be defined as:
\begin{align}
    &\epsilon_\pi(s,a) = q_\pi(s,a) - q_\pi^{\mc{D}}(s,a) \\
    &\epsilon_\pi = \textstyle\sum_s\mu_\pi(s)\sum_a\pi(a|s)|\epsilon_\pi(s,a)|
\end{align}
where $\mu_\pi$ is the state-visitation probability induced by $\pi$ in the \textit{original} MDP $\mc{M}$. It has been shown that \cite{fujimoto2019without}, $\epsilon_\pi=0$ if and only if the empirical transition probability of the batch data $\hat{p}(s^\prime|s,a)$ is equal to the true $p(s^\prime|s,a)$ for all state-action pairs $(s,a)$ with non-zero visitation probability under policy $\pi$. In this case, $q_\pi(s,a)$ can be evaluated with no error. In other words, to accurately estimate state-value functions, the agent should try to learn control policies, which tend to visit the state-action pairs contained in the batch data. A policy that satisfies this condition is denoted as batch-constrained.

\subsection{KL-Divergence Regularization and the Bellman Equation}
To find a batch-constrained policy, we propose to regularize the reward function by the \textit{KL-divergence} between the target policy and the behavior policy:
\begin{align}
r^{\m{d}}(s,a) = r(s,a) -\tau D_\m{KL}(\pi(\cdot|s)||\pi^b(\cdot|s))
\end{align}
where $r(s,a)$ is the reward function of the original MDP. $\pi^b(a|s)$ is the behavior policy, which has the same conditional probability distribution of the actions given state as that of the historical data. The KL-divergence can be calculated as $D_\m{KL}(\pi(\cdot|s)||\pi^b(\cdot|s)) = \mathbb{E}_{a\sim \pi}\left[\log\pi(a|s) - \log\pi^b(a|s)\right]$. This term encourages the agent to learn batch-constrained policies that are similar to the policy generating the historical operational data. We can rewrite the KL-divergence as 
\begin{equation}
D_\m{KL}(\pi(\cdot|s)||\pi^b(\cdot|s)) = H(\pi(\cdot|s),\pi^b(\cdot|s)) - H(\pi(\cdot|s))
\end{equation}
where $H(\pi(\cdot|s),\pi^b(\cdot|s))$ is the cross entropy of $\pi(\cdot|s)$ and $\pi^b(\cdot|s)$. $H(\pi(\cdot|s))$ is the entropy of the target policy. Therefore minimizing the KL-divergence can be thought of as maximizing the target policy's entropy coupled with minimizing the cross entropy.

We denote the value functions for a given policy $\pi$ with KL-divergence regularized reward function as $v_\pi^\m{d}(s)$ and $q_\pi^\m{d}(s,a)$. 

The Bellman equations under this setup are derived as:
\begin{align}\label{eq21}
v_\pi^{\m{d}}(s) = \mathbb{E}_{a\sim\pi}\mathbb{E}_{s^\prime\sim P}\left[ r+\gamma v_\pi^{\m{d}}(s^\prime)\right] - \tau D_\m{KL}(\pi(\cdot|s)||\pi^b(\cdot|s))
\end{align}
\begin{align}\label{eq22}
q_\pi^{\m{d}}(s,a) = r + \gamma\mathbb{E}_{s^\prime\sim P}\left[ v_\pi^{\m{d}}(s^\prime)\right]
\end{align}
\begin{align}\label{eq23}
v_\pi^{\m{d}}(s) = \mathbb{E}_{a\sim\pi}[q_\pi^{\m{d}}(s,a)] - \tau D_\m{KL}(\pi(\cdot|s)||\pi^b(\cdot|s))
\end{align}

Our next result shows that for a given policy $\pi$, the value function $q^{\m{d}}_\pi(s,a)$ can be found by the following iterative scheme:
\begin{lemma}[Batch-Constrained Soft Policy Evaluation]\label{lemma1}
	Consider the operator $\mc{T}^\pi$ given by:
	\begin{align}
	&\mc{T}^\pi q(s,a) = r(s,a) + \gamma \mathbb{E}_{s^\prime\sim P}[v(s^\prime)] \quad \forall s,a\\
	&v(s^\prime) = \mathbb{E}_{a^\prime\sim \pi}[q(s^\prime,a^\prime)] - \tau D_\m{KL}(\pi(\cdot|s^\prime)||\pi^b(\cdot|s^\prime))
	\end{align}
	and an initial $q^0(s,a)\in \mathbb{R},\forall (s,a)\in \mc{S}\times\mc{A}$. Assuming that $D_\m{KL}(\pi(\cdot|s)||\pi^b(\cdot|s))$ is bounded for all $s\in \mc{S}$, the sequence defined by $q^{k+1}=\mc{T}^\pi q^k$ will converge to the KL-divergence regularized Q function $q^\m{d}_\pi$ as $k\rightarrow \infty$.
\end{lemma}
After $q^{\m{d}}_\pi(s,a)$ is computed, we can invoke the following update rule to find an improved policy $\pi^\prime$.
\begin{lemma}[Batch-Constrained Soft Policy Improvement]\label{lemma2}
	Given a policy $\pi$ and its soft Q function $q^{\m{d}}_\pi$, define a new policy $\pi^\prime$ as follows:
	\begin{equation}
	\pi^\prime(\cdot|s) = \arg\max_{\tilde{\pi}}\; \mathbb{E}_{a\sim\tilde{\pi}}[q_\pi^{\m{d}}(s,a)] - \tau D_\m{KL}(\tilde{\pi}(\cdot|s)||\pi^b(\cdot|s)) \nonumber
	\end{equation}
	for every $s\in\mc{S}$. Then $q^\m{d}_{\pi^\prime}(s,a)\geq q^\m{d}_{\pi}(s,a)$ for all $(s,a)\in\mc{S}\times\mc{A}$.
\end{lemma}
By Lemma \ref{lemma1} and Lemma \ref{lemma2}, we can establish the following batch-constrained version of the policy iteration theorem:
\begin{theorem}[Batch-Constrained Soft Policy Iteration]\label{theorem1}
	Starting from any policy $\pi$ and alternatively applying the batch-constrained soft policy evaluation and improvement, the sequence of policies converges to a policy $\pi_*$ such that $q^\m{d}_{\pi_*}(s,a) \geq q^\m{d}_{\pi}(s,a)$ for all $(s,a)\in\mc{S}\times\mc{A}$.
\end{theorem} 
All proofs can be found in Appendix A.

Theorem \ref{theorem1} establishes the theoretical foundation for finding the optimal batch-constrained soft policy. However, it cannot be directly implemented due to infinite state space and finite training data in the dynamic DNR problem. Later in this section, we will derive a practical algorithm that approximately implements the batch-constrained soft policy iteration. Before that, we first provide an overview of the proposed reinforcement learning based dynamic DNR control framework in the next subsection.

\subsection{Overview of the Proposed Framework}
\begin{figure}[h]
	\centering
	\includegraphics[width=9cm]{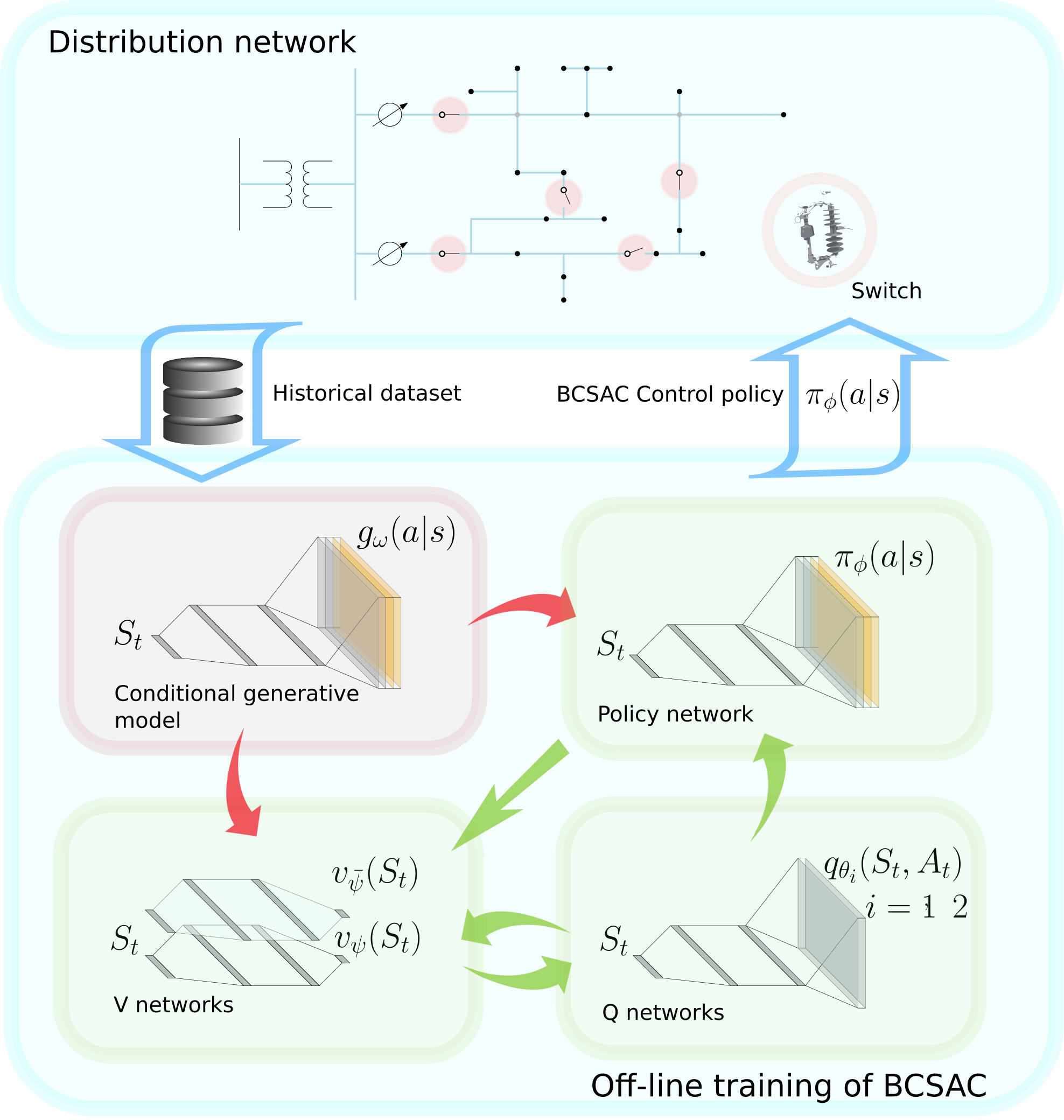}
	\caption{The proposed RL based dynamic DNR control framework}
	\label{fig_framework}
\end{figure}

This subsection provides an overview of the proposed RL based dynamic DNR control framework. Fig. \ref{fig_framework} shows the sub-modules of the proposed framework.

The electric utility first collects the historical operational dataset as described in Section II.C. This dataset will then be used for off-line training of the proposed batch-constrained soft actor-critic (BCSAC) RL algorithm. The algorithm consists of a conditional generative model, represented by the red block, and three groups of neural networks represented by the three green blocks. The conditional generative model is trained independently from the other neural networks and thus marked as red. The neural networks in the three green blocks are trained simultaneously. The red and green arrows represent the dependencies among the neural network training processes. After off-line training, the ``policy network" module will contain a trained neural network $\pi_\phi(a|s)$, which takes the network configuration and injection pattern as the input, and outputs a reconfiguration action. The policy network is trained to approximate the optimal batch-constrained soft policy $\pi_*(a|s)$. The V and Q networks are trained to approximate $v_{\pi_*}^\m{d}(s)$ and $q_{\pi_*}^\m{d}(s,a)$, respectively. In the next subsection, we present the details of the off-line training processes.

\subsection{Batch-Constrained Soft Actor-Critic}
We propose an actor-critic algorithm which approximates the policy iteration and hence, learns a batch-constrained policy from the finite historical operational dataset.
The algorithm consists of a \textit{critic}, which approximates $v_{\pi_*}^\m{d}(s)$ and $q_{\pi_*}^\m{d}(s,a)$, and an \textit{actor}, which approximates $\pi_*(a|s)$. 
\subsubsection{The Critic}
We parameterize $v_\pi^{\m{d}}(s)$ and $q_\pi^{\m{d}}(s,a)$ by neural networks and update them using the sample estimate of RHS of the equations (\ref{eq22}-\ref{eq23}). In addition, we adopt the target value network \cite{mnih2015human} and the clipped-double Q method \cite{fujimoto2018addressing} to stabilize the training. Specifically, we maintain four neural networks $q_{\theta_1}, q_{\theta_2}, v_\psi, v_{\bar{\psi}}$, and update them by:
\begin{align}\label{trainQ}
	\hspace*{-0.35cm} \min_{\theta_i} \; \frac{1}{|\mc{B}|}\sum_{(s,a,r,s^\prime)\in \mc{B}} \left[q_{\theta_i}(s,a) - (r+\gamma v_{\bar{\psi}}(s^\prime))\right]^2\; i=1,2
\end{align}
\begin{align}\label{trainV}
	\min_{\psi} \; (1/|\mc{B}|)\sum_{(s,a,r,s^\prime)\in \mc{B}} (v_\psi(s) -  v^{\m{target}}(s))^2
\end{align}
\begin{align}\label{Vtar}
    v^{\m{target}}(s) = \min_{i=1,2}q_{\theta_i}(s,\hat{a})- \tau \m{log}(\pi_\phi(\hat{a}|s)) + \tau \m{log}(\pi^b(\hat{a}|s))
\end{align}
\begin{align}\label{trainVtar}
    \bar{\psi} \leftarrow \rho \bar{\psi} + (1-\rho) \psi 
\end{align}
where $\mc{B}$ is a mini-batch sampled from the historical data $\mc{D} = \{(s,a,r,s^\prime)\}$. $\rho$ is an exponential smoothing parameter. $\hat{a}$ is a sampled action from the policy network $\pi_\phi(\cdot|s)$. The input of all the neural networks is the state $s$. For the value networks $v_\psi$ and $v_{\bar{\psi}}$, the output is a single number indicating the state value. The output of the Q networks $q_{\theta_1}$ and $q_{\theta_2}$ is a vector including the action-values. All networks are standard feedforward neural networks with a number of hidden layers. For the dynamic DNR problem, the detailed architecture design of the Q and V networks are described in Section III.H.

When performing the minimization (\ref{trainQ}) to train the Q networks $q_{\theta_1}$ and $q_{\theta_2}$, the parameter vector $\bar{\psi}$ is held fixed. Similarly, when performing the minimization (\ref{trainV}), all the parameters appearing in $v^{\m{target}}(s)$ are fixed and only $\psi$ is to be optimized. The training data of these networks are obtained from the historical operational dataset and converted into the state $s$, action $a$, reward $r$, and next state $s^\prime$ format. The process was described in Section II.B-C. In addition, sampled actions $\hat{a}$ from the current policy are also used for the training. But these samples $\hat{a}$ do not need to be the same as the actions in the historical dataset.

Next, we discuss the design of the actor and the derivation of the policy gradient.
\subsubsection{The Actor}
We approximate the policy function (the actor) by a neural network parameterized by $\phi$. 
Ideally, the parameters should be updated using gradient ascent
$\phi \leftarrow \phi + \eta \nabla v_{\pi_\phi}^{\m{d}}(s)$, where the $\nabla v_{\pi_\phi}^{\m{d}}(s)$ is given by
\begin{align}
    &\nabla v_{\pi_\phi}^{\m{d}}(s) 
     = \nabla [\mathbb{E}_{a\sim\pi_\phi}[q_\pi^{\m{d}}(s,a)] - \tau D_\m{KL}(\pi_\phi(\cdot|s)||\pi^b(\cdot|s))] \nonumber \\
    & = \nabla \mathbb{E}_{a\sim\pi_\phi}[q_\pi^{\m{d}}(s,a) - \tau (\log\pi_\phi(a|s) - \log\pi^b(a|s))] \label{eq26}
\end{align}
However, this policy gradient requires computing the derivative of $q_\pi^{\m{d}}(s,a)$. It is shown that this derivative will result in an \textit{on-policy} policy gradient \cite{sutton2018reinforcement}, which cannot be estimated from a given historical dataset. 

Fortunately, as shown by Lemma 2, the gradient of $q_\pi^{\m{d}}(s,a)$ can be omitted. This is because the objective function of Lemma 2 treats $q_\pi^{\m{d}}(s,a)$ as a constant. In other words, updating the actor without the gradient information of $q_\pi^{\m{d}}(s,a)$ still approximates the monotonic policy improvement. With this theoretical guarantee, we can derive an \textit{off-policy} policy gradient, which can be estimated from the historical dataset. The derivation is done in three steps. In the first step, we omit the gradient of $q_\pi^{\m{d}}(s,a)$.
This is justified by Lemma 2. In the second step, we change the order of the gradient operator and the expectation operator and define a new term $f_\phi(s,a) = q_\pi^{\m{d}}(s,a) - \tau (\log\pi_\phi(a|s) - \log\pi^b(a|s))$ to simplify the notation.
\begin{align}
    &\nabla \mathbb{E}_{a\sim\pi_\phi}[f_\phi(s,a)] \nonumber \\
    =&\sum_a f_\phi(s,a) \nabla \pi_\phi(a|s) - \tau \underbrace{\sum_a \pi_\phi(a|s)\nabla \log \pi_\phi(a|s)}_{=0} \nonumber \\
    =&\sum_a \pi_\phi(a|s) f_\phi(s,a) \nabla \log\pi_\phi(a|s) \nonumber \\
    =&\mathbb{E}_{a\sim\pi_\phi} f_\phi(s,a)\nabla \log \pi_\phi(a|s)\label{step2}
\end{align}
where we have used the identity $\sum_a\pi_\phi(a|s)\nabla \log\pi_\phi(a|s) = \sum_a\nabla\pi_\phi(a|s) = \nabla\sum_a\pi_\phi(a|s)=\nabla 1 = 0$. In the third step, we replace the expectation in (\ref{step2}) by its one-sample estimate. The final form of the approximate policy gradient is given by:

\begin{align}\label{gradpi}
    \hat{\nabla} v_{\pi_\phi}^{\m{d}}(s) = \nabla\log &\pi_\phi(\hat{a}|s)[q_{\theta_1}(s,\hat{a}) \nonumber \\
    &- \tau(\log\pi_\phi(\hat{a}|s) - \log \pi^b(\hat{a}|s))]
\end{align}
where $\hat{a}$ is sampled from $\pi_\phi(\cdot|s)$. This completes the derivation of the actor network update process.

The structure of the policy network is as follows. The input of $\pi_\phi$ is the state $s$, and the output is a conditional probability distribution of actions given the state. For the dynamic DNR problem, the detailed architecture for the policy network is described in Section III.H. Note that in order to evaluate \eqref{Vtar} and \eqref{gradpi}, we need to approximate the behavior policy $\pi^b(\hat{a}|s)$ by a parametric function $g_\omega(\hat{a}|s)$. This will be discussed in the next subsection.

\subsection{Representing the Batch Distribution as a Parametric Model} \label{CVAE}
Given the difficulty of estimating the behavior policy $\pi^b(a|s)$ with high-dimensional state and action space, we propose using the conditional variational autoencoder (CVAE) \cmmnt{\cite{doersch2016tutorial}} \cite{Sohn2015generative} as the parametric generative model for $g_\omega(a|s)$, where $\omega$ is the model parameter. Using non-parametric models for this learning task can be very difficult because they suffer from the curse of dimensionality. Furthermore, non-parametric models have difficulty handling mixed discrete and continuous variables. On the other hand, CVAE model is well suited for our application due to three reasons. First, CVAE model is very scalable and can approximate high-dimensional distributions. Second, CVAE model can easily handle mixed discrete and continuous state-action space. Third, as will be shown in Section IV.C, CVAE model has good empirical performance for our application.

CVAE consists of an encoder $c_{\omega^\prime}(z|s,a)$, which maps a given state-action pair to a latent representation $z$, and a decoder $g_\omega(a|s,z)$, which produces the probability of taking an action $a$ given $z$ and $s$. CVAE maximizes the following objective function to obtain the parameters for the encoder, $\omega^\prime$, and the decoder $\omega$:
\begin{align}
\mathbb{E}_{z\sim d_{\omega^\prime}}[\log g_\omega(a|s,z)] - D_\m{KL}(c_{\omega^\prime}(z|s,a)||p(z))
\end{align}
where $p(z)$ is the latent variable distribution and is chosen as a Gaussian $\mc{N}(0,I)$. To train CVAE, we sample mini-batches of state-action pairs $(s,a)$ from the historical data $\mc{D}$ and perform stochastic gradient ascent for the sample objective function. The trained decoder $g_\omega(a|s,z)$ is used to represent the behavior policy of the historical operational data.

\subsection{Summary of BCSAC Algorithm}
Our proposed batch-constrained soft actor-critic (BCSAC) algorithm is summarized in Algorithm \ref{algo2}. The algorithm takes as inputs the operational historical dataset $\mc{D}$ (the batch) as well as the trained CVAE model $g_\omega$. Before the training starts, the policy and value networks are initialized using general-purpose deep neural network initialization algorithms (the Xavier initialization). In each iteration, the algorithm first samples a mini-batch of experiences from the batch, and then samples actions from the current policy. Afterwards, the algorithm conducts policy evaluation by training the V and Q-networks using (\ref{trainV}) and (\ref{trainQ}), respectively. At the end of each iteration, the policy improvement step is taken by training the policy network, which updates the parameters $\phi$ using the gradient shown in (\ref{gradpi}). Note that since the historical dataset $\mc{D}$ does not change during the training process, the trained CVAE model $g_\omega$ does not need to be updated.

The proposed algorithm differs from existing actor-critic frameworks (e.g. \cite{haarnoja2018soft}) in three ways. First, the framework is developed from a novel batch-constrained soft policy iteration theory presented in Theorem \ref{theorem1}. Second, we utilize finite action space policy gradient in (\ref{gradpi}) to update the actor network, instead of the reparameterization trick \cite{haarnoja2018soft}. Third, a pre-trained conditional generative model is incorporated for the training of the batch-constrained RL algorithm.

For the dynamic DNR problem, the historical data $\mc{D}$ consists of the nodal power injections, substation SCADA power measurements, nodal voltage magnitudes, and the status of remotely controllable switches. These data have been converted into the state, action, reward, next state tuple $(s,a,r,s^\prime)$ prior to the training. The detailed procedure was described in Section II.B-II.C. To apply the BCSAC algorithm to the dynamic DNR problem, we design unique neural network architectures and the representation of distribution network topology in these networks. This is the subject of the next subsection.
\begin{algorithm}[h]
	\caption{BCSAC with Finite Action Space}
	\label{algo2}
	\textbf{Input:} Batch $\mc{D}$, conditional generative model $g_\omega \approx \pi^b$
	\begin{algorithmic}[1]
		\State Initialize $\phi, \theta_1, \theta_2, \psi, \bar{\psi}$
		\For{$i=1,\cdots,$}{}
		\State Sample mini-batch $\mc{B}=\{(s,a,r,s^\prime)\}$ from $\mc{D}$
		\State Sample actions from the current policy: $\hat{a}\sim \pi_\phi(\cdot|s)$
		\State Train Q networks $\theta_1, \theta_2$ by (\ref{trainQ})
		\State Train V network $\psi$ by (\ref{trainV})
		\State Update V target network $\bar{\psi}$ by (\ref{trainVtar})
		\State Train policy network $\phi$ by $\phi \leftarrow \phi + \eta \hat{\nabla} v_{\pi_\phi}^{\m{d}}(s)$ 
		\Statex[1] where $\hat{\nabla} v_{\pi_\phi}^{\m{d}}(s)$ is given by (\ref{gradpi})
		\EndFor
	\end{algorithmic}
\end{algorithm}

\subsection{Algorithm Implementation}
This subsection provides the technical details of implementing BCSAC algorithm for the dynamic DNR problem. The neural network architecture design and representation of distribution network topology are covered.

\noindent$\bullet$ Representation of distribution network configuration as an input to neural networks: we use a binary vector of on/off status of each line segment to encode the distribution network configurations. Since the configuration at each time step must be radial, the next feasible state configurations $\vb{\alpha}_{t+1}$ starting from an existing configuration $\vb{\alpha}_{t}$ are discovered as follows. First, we identify all closeable switches in $\vb{\alpha}_{t}$. Closing any one of these closeable switches $i$ creates exactly one fundamental cycle. Each line segment $j$ in this fundamental cycle can be opened. We store all such switchable pairs $(i,j)$ at time $t$ in a binary 2-D array $M^t$. $M^t_{ij}=1$ if $(i,j)$ is a valid switching pair, and is $0$ otherwise.

\noindent$\bullet$ Policy network $\pi_\phi$ structure: The output of the policy network is a 2-D array $\pi_{ij}(S_t)$ and is the probability distribution of switching pairs of branches $(i,j)$, that is, $\pi_{ij}(S_t)\geq 0$ and $\sum_{i=1}^m\sum_{j=1}^m\pi_{ij}(S_t)=1$. $\pi_{ij}(S_t)$ must be zero if $M^t_{ij}=0$. To enforce this, we use a masked softmax layer as the output of the policy network: 
\begin{align}
\pi_{ij}(S_t) = \frac{e^{h_{ij}(S_t)} \cdot M^t_{ij}}{ \sum_{kl} e^{h_{kl}(S_t)} \cdot M^t_{kl}}
\end{align}
where $h_{ij}(S_t)$ are the outputs of the previous layer. $M^t_{ij}$ is the binary mask. The same masked softmax layer is used as the output layer of the parametric generative model $g_\omega$.

\noindent$\bullet$ Q-network $q_\theta$ structure: The input to the Q-network is the state encoding, along with the one-hot encoding of the closeable switches. The number of outputs of the Q-network equals the number of switches of the distribution network, which correspond to openable switches.

\noindent$\bullet$ V-network $v_\psi$ structure: The value network $v_\psi(S_t)$ is a standard multilayer perceptron. The input of the V-network is the state encoding and the output is the value of that state.

\section{Numerical Studies}
To verify the performance of our proposed BCSAC algorithm on dynamic DNR problems, we conduct comprehensive numerical studies on four distribution networks. We start by presenting the experimental data and the algorithm setup in Subsections IV.A-IV.B. The optimality, scalability, and computation efficiency of the proposed algorithm and benchmark algorithms are shown in Subsections IV.C-IV.F.

\subsection{Experimental Data Setup}
\subsubsection{Distribution Networks}
\begin{figure}[h]
	\centering
	\includegraphics[width=8cm]{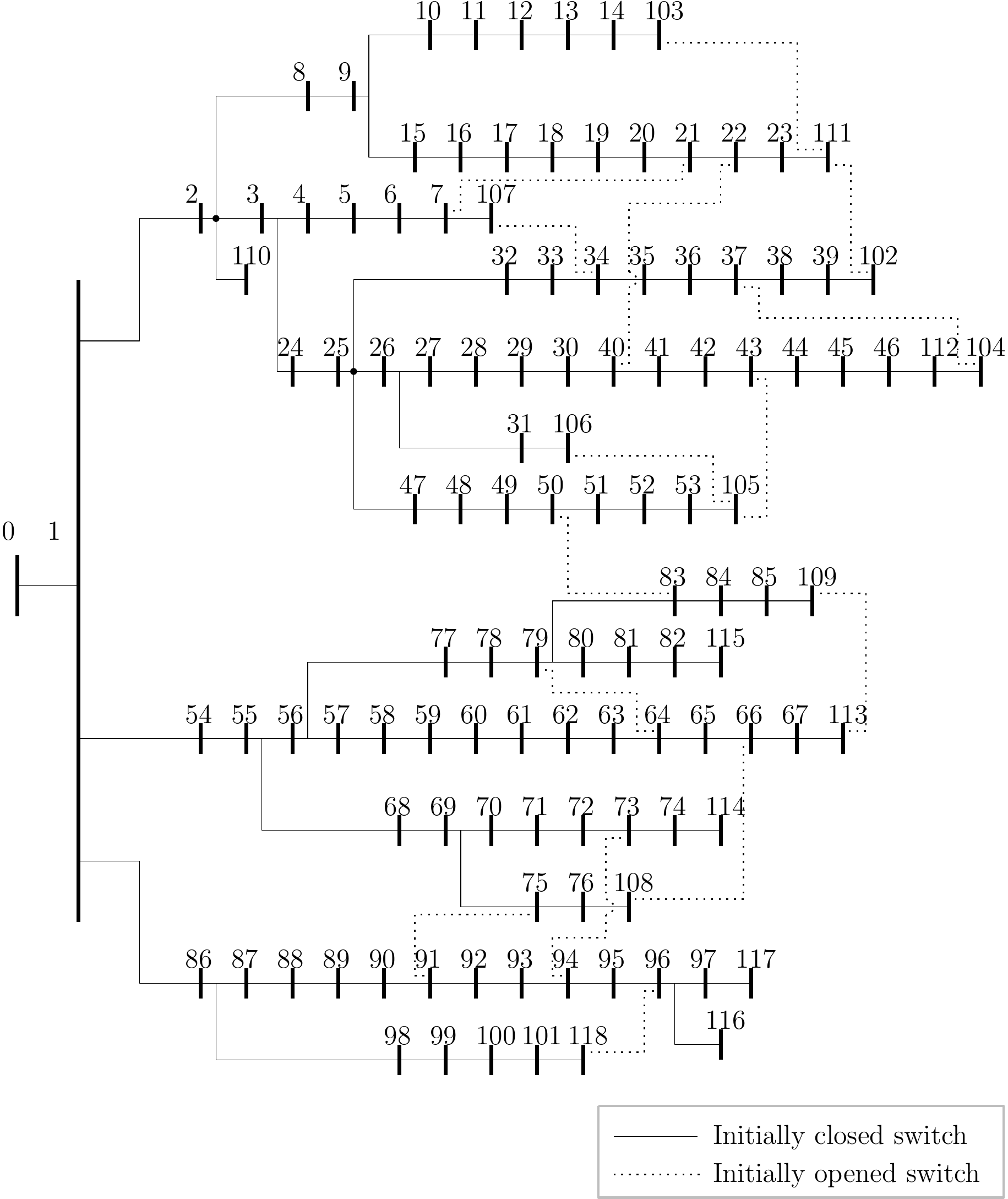}
	\caption{The 119-bus test feeder and its initial configuration}
	\label{fig119bus}
\end{figure}

The 16-bus \cite{su2003network}, 33-bus \cite{Baran1989reconfig}, 70-bus \cite{Das2006fuzzy}, and 119-bus \cite{zhang2007improved} distribution networks are chosen for the numerical study. The schematic diagram of the 119-bus distribution network is shown in Fig. \ref{fig119bus}. For notational convenience, we have modified the bus numbering described in \cite{zhang2007improved}. It is assumed that each line segment has a remotely controllable switch. The total number of feasible configurations is used as a measure of complexity of the learning task and is shown in Table \ref{table1}. The number of feasible configurations are calculated by matrix-tree theorem \cite{Gao1910:Dynamic}. Note that the number of feasible configurations increases exponentially with the number of remotely controllable switches. In Table \ref{table1}, the Solar bus column shows the buses with solar generation. For all test cases, the retail electricity price $C^l$ is set as 0.13 \$/kWh. The maximum and minimum nodal voltages are set as $\bar{v}=1.1$ and $\underline{v}=0.9$, and the voltage violation penalty is set as $\lambda=C^l$. An alternative modeling approach is to use hard constraints to limit the variations of voltage. For example, constrained policy optimization \cite{achiam2017constrained} and constrained soft actor-critic \cite{Wei2019Safe} can be implemented to eliminate the need to specify $\lambda$. However, these methods are either on-policy or require implementing several additional neural networks. As such, we propose selecting $\lambda$ based on operational considerations and empirical performance. The switching cost $C^s$ that appeared in (\ref{eq3}) is also given in Table \ref{table1}.

\begin{table}[h]
\setlength{\tabcolsep}{1pt}
    \centering
	\caption{Test Distribution Networks}
	\begin{tabular}{C{1.1cm}C{0.9cm}C{3cm}C{0.9cm}R{2.7cm}}
		\toprule  
		Case & $S_{base}$ (MVA) & Solar buses & $C^s$ (\$) & \# configuration \\ \hline
	    16-bus & 100 & \{11\} & 4.0 & 190  \\
		33-bus & 175 & \{4,6,12\} & 0.5 & 50,751\\
		70-bus & 500 & \{8,10,26,28,50,52\} & 2.0 &  22,621,020,015 \\ 
		119-bus & 500 & \{33,45,46,55,80,86,101\} & 0.8 & 3,853,525,605,824,176 \\ 
		\bottomrule
	\end{tabular}
	\label{table1}
\end{table}

\subsubsection{Nodal Power Data}
The time series of load data are taken from the Irish Commission for Energy Regulation Smart Metering Project \cite{IrishDataset}. The dataset contains one and a half years (76 weeks) of smart meter kWh measurements from approximately 1,000 customers. For each of the test distribution networks, we aggregate the power consumption from 30 (15 for the 70-bus and 119-bus network) customers as the nodal real power injections. We assume a constant power factor of 0.98 lagging. For each of the test networks, the solar generation data are obtained from southern California sites \cite{2015solar}. All nodal power injections are scaled by a common factor $\beta$ (i.e., $(\vb{p}_t,\vb{q}_t) \mapsto (\beta \vb{p}_t,\beta \vb{q}_t)$ for all $t$) to create a realistic network loading level. $\beta$ is chosen such that the resulting average total line losses are roughly $1.5\%$ of the total demand \cite{Gao1910:Dynamic}. For all case studies, the first 52 weeks of data are used for training and data of the following week are used for testing. 

\subsubsection{Network Configuration Data}
The last piece of information in the historical operational data is the network configuration data. Unfortunately, we are unable to obtain real world switch configuration data. Thus, the historical configuration data is created by the simulation. In practice, the algorithm will be trained on real world data rather than simulated ones. Therefore, no network parameter information is needed. We create different sets of historical configuration data as follows. At each time step $t$, the configuration $\vb{\alpha}_{t}$ can be changed to $\vb{\alpha}_{t+1}$ by a single pair of branch-exchange in one of the three scenarios:
\begin{enumerate}
    \item[s.1] The network is reconfigured by the one-step model-based reconfiguration algorithm assuming inaccurate knowledge of network parameters. To simulate a model-based controller with inaccurate information, we synthesize a different set of line parameters, which deviate from their true values by 10\%. We used the mixed-integer conic programming (MICP) formulation in \cite{Dorostkar2016Value} with a time horizon of 1 hour and the number of switching actions of 2 per time step. 
    \item[s.2] The network configuration is kept the same.
    \item[s.3] The configuration is randomly changed to another constraint-satisfying topology.
\end{enumerate}

Scenario 1 represents the active distribution grid reconfiguration performed by a model-based controller with inaccurate information. Scenario 2 corresponds to passive grid management or periods with SCADA system failure, where the network configuration stays the same. Scenario 3 represents periods with isolating faults, when network reconfiguration must be performed to restore power. To create a synthetic network reconfiguration sequence, at time $t$, we choose a scenario to obtain the new network configuration based on the probability assigned to each scenario. We denote probabilities for the three scenarios as $P_{mod}$, $P_{fix}$, and $P_{rnd}$. By varying these three probabilities, we obtain historical dataset for network configurations with different characteristics. In particular, $P_{mod}=1$ corresponds to the case where a model-based controller with inaccurate network parameter information is always used to reconfigure the distribution network. The initial configurations $\vb{\alpha}_0$ of all datasets are the all-tie-switch-open configuration.

\subsection{Algorithm Setup}
The setup of the proposed BCSAC algorithm and two benchmark RL algorithms are summarized in this subsection. The hyperparameters of the BCSAC algorithm and the benchmark DQN and SAC algorithms are provided in Table \ref{table2}. The hyperparameters of the three RL algorithms are tuned individually to reach their best performance. The last row of Table \ref{table2} shows the parameters shared by all algorithms. Note that we scale the reward (in per unit) to match the weights of neural networks. If not specified otherwise, these parameters will be used for all the numerical studies. Four parameters in the curly brackets are for the three distribution networks, from left to right, 16, 33, 70, and 119-bus, respectively. We also compare the performance of the proposed BCSAC algorithm with that of the historical operational strategy, which is a mix of the model-based, passive, and random control scenarios.

\begin{table}[h]
	\caption{Hyperparameters of RL Algorithms}
	\begin{tabular}{l p{3cm} p{3.6cm}}
		\toprule  
		\multirow{1}{*}{DQN} & learning rate & $\{10^{-4},10^{-4},10^{-4},10^{-4} \}$  \\ 
		    &number of hidden units & $200, 200, 250, 250 \}$ \\
		    &copy steps & $\{30,30,30,30 \}$ \\ 
		    & minibatch size & $\{32,64,64,64\}$ \\\hline
		\multirow{2}{*}{SAC} & $\tau$ & $\{0.002,0.001,0.0005,0.0005\}$  \\ 
		    & learning rate & $\{5\cdot 10^{-4},10^{-4},10^{-4},10^{-4}\}$  \\ 
		    & number of hidden units & $100, 200, 200, 250\}$ \\
		    & $\rho$ & $\{0.99,0.99,0.99,0.99\}$ \\ 
		    & minibatch size & $\{32,64,64,64\}$ \\\hline    
		\multirow{2}{*}{BCSAC} & $\tau$ & $\{0.1,10,25,50\}$  \\ 
		    & learning rate & $\{10^{-4},10^{-4},5\cdot 10^{-5},5\cdot 10^{-5}\}$  \\ 
		    & number of hidden units & $\{100,100,200,250\}$ \\
		    & $\rho$ & $\{0.995,0.995,0.995,0.995\}$ \\ 
		    & minibatch size & $\{32,32,64,64\}$ \\\hline        
		\multirow{2}{*}{CVAE} 
		    & learning rate & $10^{-4}$  \\ 
		    & number of hidden units & 1400 \\
		    & latent space dimension & $\{20,40,60,70\}$ \\ \hline        
		\multirow{2}{*}{shared} &discount factor & $0.95$  \\
		       &number of hidden layers & 2 \\
		       &hidden unit nonlinearity & ReLU  \\
		       &optimizer & Adam \\
		       &reward scale & 500 \\
		\bottomrule
	\end{tabular}
	\label{table2}
\end{table}

\subsection{Approximating Behavior Policy by CVAE}
\begin{figure}[h]
	\centering
	\includegraphics[width=8cm]{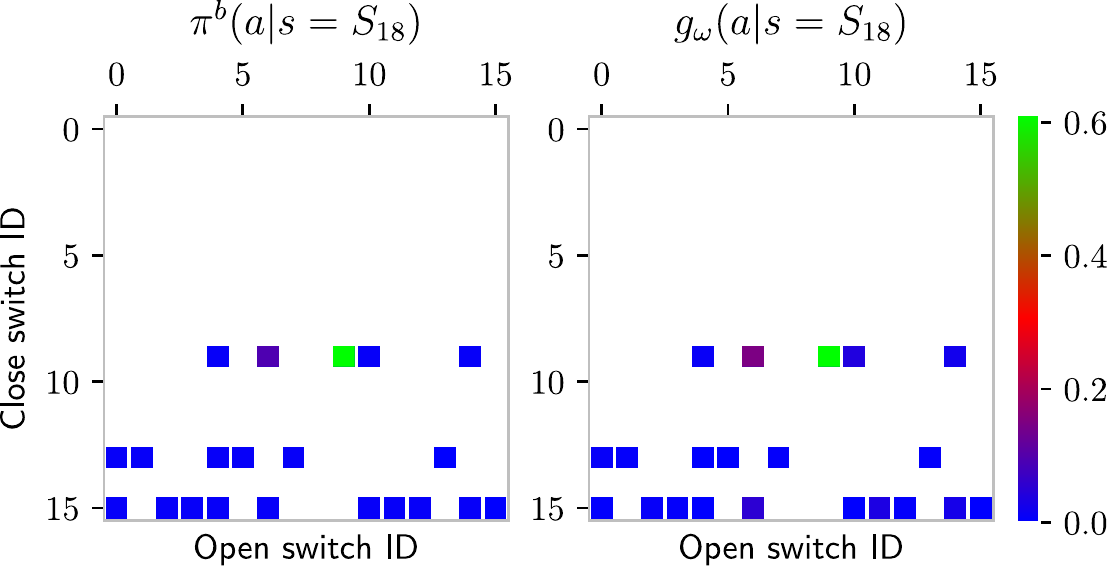}
	\caption{A sample result of CVAE on the 16-bus test feeder (TV-distance $=0.19$)}
	\label{fig_cvae}
\end{figure}

This subsection provides the experimental justification of using the CVAE model $g_\omega(a|s)$ to approximate the behavior policy $\pi^b(a|s)$. We first present the performance of CVAE on one of the synthetic datasets. The sample synthetic dataset is obtained with $[P_{mod}, P_{fix}, P_{rnd}]=[0.1, 0.72, 0.18]$ for the 16-bus feeder. We train the CVAE model to approximate the behavior policy. Fig. \ref{fig_cvae} shows the ground-truth $\pi^b(a|s=S_{18})$ and the CVAE approximation $g_{\omega}(a|s=S_{18})$ for the 18-th time step of the dataset. In Fig. \ref{fig_cvae}, the $(i,j)$-th cell of each of the table shows the discrete probability of closing switch $i$ and opening switch $j$. The cell with the highest probability corresponds to fixing the configuration (Scenario s.2); the cell with the second largest probability corresponds to the branch-exchange obtained from MICP (Scenario s.1); the other cells correspond to randomly changing reconfiguration (Scenario s.3). Cells correspond to infeasible opening/closing pairs (result in non-radial configuration) are left as white.

Fig. \ref{fig_cvae} shows that, the CVAE model approximates the behavior policy for all reconfiguration actions $a$ at state $S_{18}$ with high accuracies, \textit{even if the training dataset only contains one reconfiguration action at this state}. In other words, the trained CVAE model can generalize the training dataset to unseen state-action pairs.

We use the total variation (TV) distance between $\pi^b(a|s)$ and $g_{\omega}(a|s)$ to measure their dissimilarity: 
\begin{equation*}
||\pi^b(\cdot|s) - g_{\omega}(\cdot|s)||_{\m{TV}} = \frac{1}{2}\sum_{a\in\mc{A}(s)}|\pi^b(a|s)-g_\omega(a|s)|
\end{equation*}

In Table \ref{tableCVAE}, we report the average TV distance across all states contained in $\mc{D}$. Table \ref{tableCVAE} shows that the CVAE model generalizes very well across all states and different data distributions. This makes the CVAE model well suited for training the BCSAC algorithm. This is because during the training process, different actions $\hat{a}$ might be sampled from the policy network (Algorithm \ref{algo2}, Line 4). The CVAE model always yields a good approximation for $\pi^b(\hat{a}|s)$.
\begin{table}[h]
	\caption{Average TV-Distance Between $\pi^b(a|s)$ and $g_{\omega}(a|s)$}
	\begin{tabular}{r C{0.8cm} C{0.8cm} C{0.8cm} C{0.8cm} C{0.8cm} C{0.8cm} }
		\toprule  
		$P_{mod}$ & 0.1 & 0.3 & 0.5 & 0.7 & 0.9 & 1.0  \\ \hline
		16-bus & 0.13 & 0.15 & 0.13 & 0.11 & 0.06 & 0.03 \\ 
		33-bus & 0.24 & 0.24 & 0.19 & 0.15 & 0.12 & 0.08  \\ 
		70-bus & 0.28 & 0.26 & 0.21 & 0.16 & 0.08 & 0.07 \\
	    119-bus & 0.31 & 0.24 & 0.17 & 0.10 & 0.04 & 0.01  \\ 
		\bottomrule
	\end{tabular}
	\label{tableCVAE}
\end{table}

\subsection{Optimality and Scalability}
\begin{figure}[h]
	\centering
	\includegraphics[width=8.9cm,height=4cm]{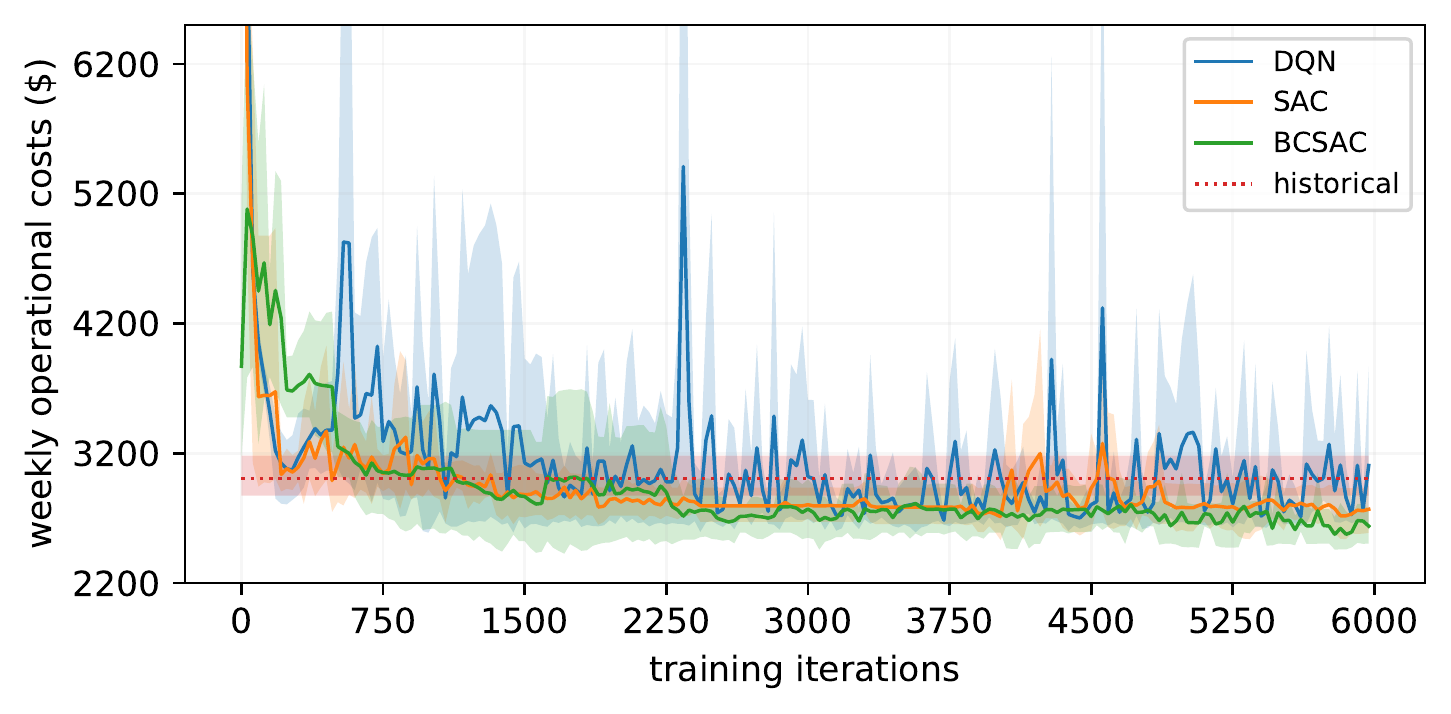}
	\caption{16-bus test feeder}
	\label{fig_16busOptimality}
\end{figure}

We first present the performance of various algorithms on one of the synthetic datasets. More comprehensive evaluations will be provided shortly. The sample synthetic network configuration dataset is obtained with $[P_{mod}, P_{fix}, P_{rnd}]=[0.5, 0.4, 0.1]$ for the 16-bus feeder. During the training process of the RL algorithms, we periodically record the weights of the value and policy neural networks, which are used to evaluate the algorithm performance on the testing week. Experiments with five random historical dataset and neural network initialization and training are conducted. Fig. \ref{fig_16busOptimality} shows the cumulative operational cost plus voltage violation penalty over the testing week.

As shown in Fig. \ref{fig_16busOptimality}, by adopting the proposed BCSAC algorithm, the RL agent is capable of finding a control policy, which yields a lower weekly operational cost than state-of-the-art RL algorithms (DQN and SAC) and historical operational strategy. It should be noted that overfitting could occur in batch RL. This is because the agent is learning from a fixed dataset rather than interacting with the environment. Nevertheless, by using a small-sized neural network and stop training early, we found both of the benchmark and the proposed BCSAC algorithm have little or no overfitting problem as shown in Fig. \ref{fig_16busOptimality}.

\begin{figure}[h]
	\centering
	\includegraphics[width=8.9cm,height=4cm]{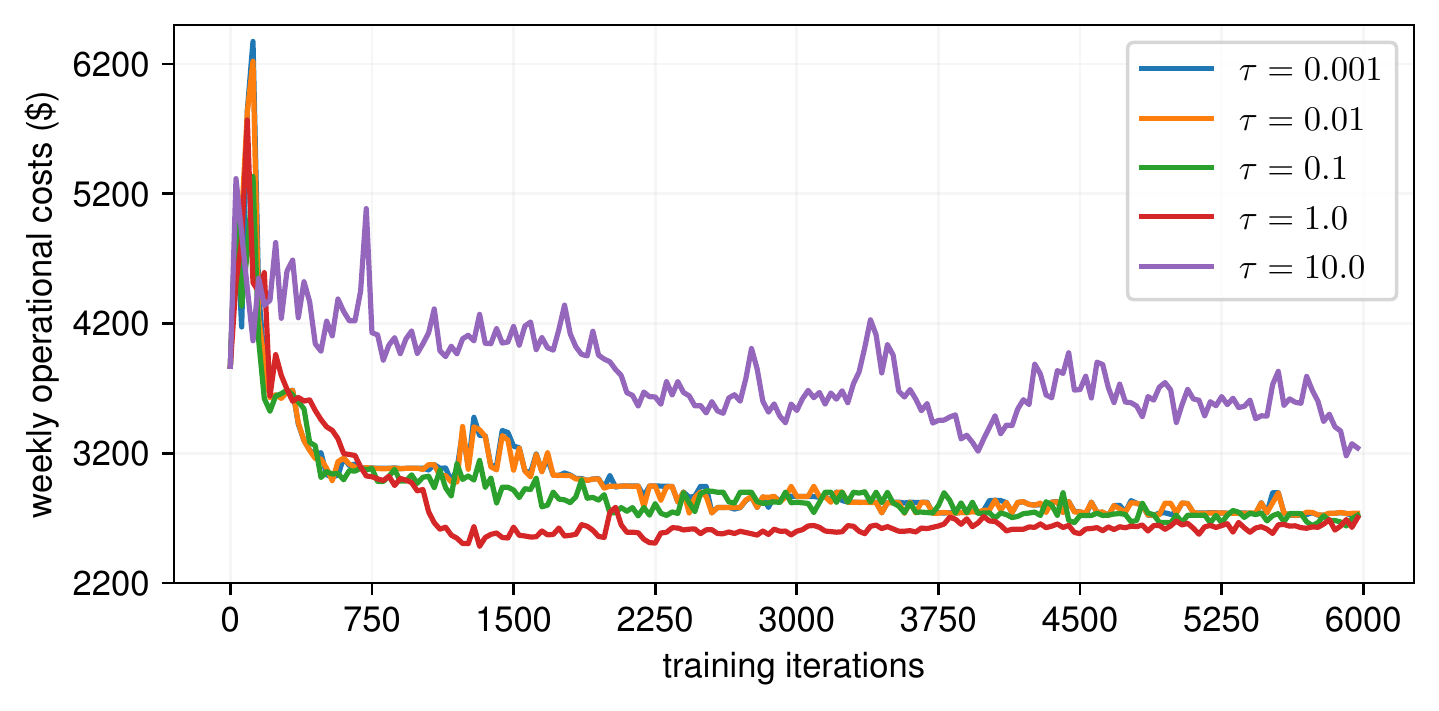}
	\caption{Sensitivity of BCSAC to the temperature parameter $\tau$ on the 16-bus test feeder}
	\label{fig_sensitivity}
\end{figure}

The selection of temperature parameter $\tau$ is very important to the BCSAC algorithm. Next, we provide a sensitivity analysis of $\tau$. Consider the same experiment as in Fig. \ref{fig_16busOptimality}, but with varying $\tau$ parameters. The median weekly operational costs of the BCSAC algorithm over 5 independent runs for 5 different temperature parameters, are shown in Fig. \ref{fig_sensitivity}.

As shown in Fig. \ref{fig_sensitivity}, the performance of the proposed algorithm does depend on the temperature parameter when $\tau$ is beyond a certain range. The performances of the algorithm are nearly identical when $\tau$ varies from $0.001$ to $0.1$. This suggests that the proposed algorithm is fairly robust with respect to the temperature parameter $\tau$. However, a very large $\tau$ does degrade the algorithm performance. This is because the policy is not learned based on the reward but mostly from the behavior policy. In practice, the value of $\tau$ should be chosen such that the numerical range of $r(s,a)$ and that of the $|\mc{A}(s)|/\tau$ are roughly within the same order of magnitude, where $|\mc{A}(s)|$ denotes the size of the action space. Both $|\mc{A}(s)|$ and $r(s,a)$ can be calculated based on the historical dataset.

Next, we conduct numerical studies on six other historical operational datasets, which are generated by varying $P_{mod}$ from 0.1 to 1.0 while fixing the ratio of $P_{fix}$ to $P_{rnd}$ at 4. Since our proposed and benchmark RL algorithms are agnostic to the data generation process, the same set of hyperparameters must be used for all test networks and datasets. The testing results on four distribution networks after 6,000 training steps are given in Table \ref{table16bus}-\ref{table119bus}. As shown in the tables, the proposed BCSAC algorithm outmatches state-of-the-art RL algorithms (DQN and SAC) for most of the experiments in terms of weekly operational costs. It consistently outperforms the RL benchmarks for large test feeders such as the 33-, 70-, and 119-bus feeders. For most of the historical datasets, the BCSAC algorithm improves the behavior policy that generates the dataset.
 
\begin{table}[h]
	\caption{Weekly operational costs for 16-bus feeder (\$)}
	\begin{tabular}{l c c c c c c }
		\toprule  
		$P_{mod}$ & 0.1 & 0.3 & 0.5 & 0.7 & 0.9 & 1.0  \\ \hline
		DQN & 3268.8 & 3036.8 & 3105.4 & 4448.3 & 2900.8 & 3604.8 \\ 
		SAC & 2832.6 & 2715.7 & 2768.3 & 3276.8 & 2595.7 & 3613.5  \\ 
		BCSAC & 2792.7 & 2642.0 & 2720.6 & 2558.4 & 2466.5 & 2451.7 \\
	    Historical & 4527.0 & 3875.2 & 3004.8 & 2685.7 & 2510.4 & 2384.7 \\ 
		\bottomrule
	\end{tabular}
	\label{table16bus}
\end{table}

\begin{table}[h]
	\caption{Weekly operational costs for 33-bus feeder (\$)}
	\begin{tabular}{l c c c c c c }
		\toprule  
		$P_{mod}$ & 0.1 & 0.3 & 0.5 & 0.7 & 0.9 & 1.0  \\ \hline
		DQN & 3613.0 & 7286.3 & 3074.7 & 3502.2 & 3211.1 & 6185.6  \\ 
		SAC & 4976.0 & 2796.3 & 2195.4 & 2250.1 & 3658.7 & 7359.2  \\ 
		BCSAC & 2388.6 & 1921.3 & 1732.3 & 1732.3 & 1716.7 & 1690.1 \\
	    Historical & 3534.6 & 2580.0  & 1961.3 & 1757.8 & 1776.4 & 1686.4 \\ 
		\bottomrule
	\end{tabular}
	\label{table33bus}
\end{table}

\begin{table}[h]
	\caption{Weekly operational costs for 70-bus feeder (\$)}
	\begin{tabular}{l c c c c c c }
		\toprule  
		$P_{mod}$ & 0.1 & 0.3 & 0.5 & 0.7 & 0.9 & 1.0  \\ \hline
		DQN & - & 6839.6 & - & - & - & - \\ 
		SAC & - & 6801.2 & 3930.0 & 5522.8 & - & 4034.6  \\ 
		BCSAC & 4143.0 & 3535.2 & 3622.1 & 3331.3 & 3449.5 & 3369.3 \\
	    Historical & 6262.0 & 4643.6 & 4437.7 & 3507.2 & 3453.6 & 3334.4 \\ 
		\bottomrule
	\end{tabular}
	\label{table70bus}
\end{table}

\begin{table}[h]
	\caption{Weekly operational costs for 119-bus feeder (\$)}
	\begin{tabular}{l c c c c c c }
		\toprule  
		$P_{mod}$ & 0.1 & 0.3 & 0.5 & 0.7 & 0.9 & 1.0  \\ \hline
		DQN & 4952.4 & 3432.3 & 7098.4 & 2881.4 & 3728.1 & 5336.5 \\ 
		SAC & 3930.0 & 3102.2 & 2715.5 & 2388.0 & 2743.9 & 4448.6  \\ 
		BCSAC & 3673.1 & 2432.3 & 2071.6 & 2102.9 & 2164.4 & 2046.2 \\
	    Historical & 4758.6 & 2811.8 & 2323.2 & 2112.7 & 2127.0 & 2046.2 \\ 
		\bottomrule
	\end{tabular}
	\label{table119bus}
\end{table}

The scalability of our proposed BCSAC algorithm is demonstrated by its performance shown in Table \ref{table70bus}-\ref{table119bus} on the 70- and 119-bus distribution network, which has more than 3.8 quadrillion feasible configurations. Learning a control strategy with limited historical operational data is extremely difficult on these test cases. This is because, the historical operational data only contain an extremely small subset of all feasible state-action pairs. 
The DQN and SAC algorithm even fail to learn a dynamic DNR strategy for the highly resistive 70-bus feeder \cite{Das2006fuzzy}. On the other hand, by learning a batch-constrained policy, our proposed BCSAC algorithm not only outperforms DQN and SAC, but also outmatches the existing behavior control policy, which is a mixed model-based, passive, and random control strategy.

\subsection{Behavior of BCSAC in Response to Unforeseen States}
In this subsection, we test how the trained BCSAC agent would respond to an unforeseen/extreme scenario during the testing time. These scenarios are likely to happen in actual grid operation. For example, power injection patterns might change abruptly due to extreme weather condition or special event. the RL control policy might be overridden by a human operator to perform higher priority tasks such as fault isolation, resulting in an ``unfamiliar" network configuration to the RL agent. In any case, the RL agent is expected to perform safely and effectively. 

Consider the 16-bus feeder and the same training dataset $[P_{mod}, P_{fix}, P_{rnd}]=[0.7, 0.24, 0.06]$. We train the BCSAC algorithm by the same procedure as described in Section IV.D. This trained BCSAC agent is then evaluated on two experiments.

$\bullet$ In the first experiment, we intentionally change the network configuration to some new configurations $\tilde{\bm{\alpha}}_t$ for each hour $t$ of the testing week. Each of these configurations does not appear in the historical dataset. 

$\bullet$ In the second experiment, we change the power injections to some extreme patterns $[\tilde{\vb{p}}_t,\tilde{\vb{q}}_t]$ for each hour of the testing week. These patterns are obtained by disconnecting the solar generation and connecting large amount of loads at various buses. The resulting injection patterns deviate significantly from the training dataset's average $[\bar{\vb{p}},\bar{\vb{q}}]$. That is, $||[\tilde{\vb{p}}_t,\tilde{\vb{q}}_t]-[\bar{\vb{p}},\bar{\vb{q}}]||_2$ is 1.3 times greater than the largest deviation within the training dataset for a typical $t$. 

\begin{figure}[H]
	\centering
	\includegraphics[width=7.5cm, height=5cm]{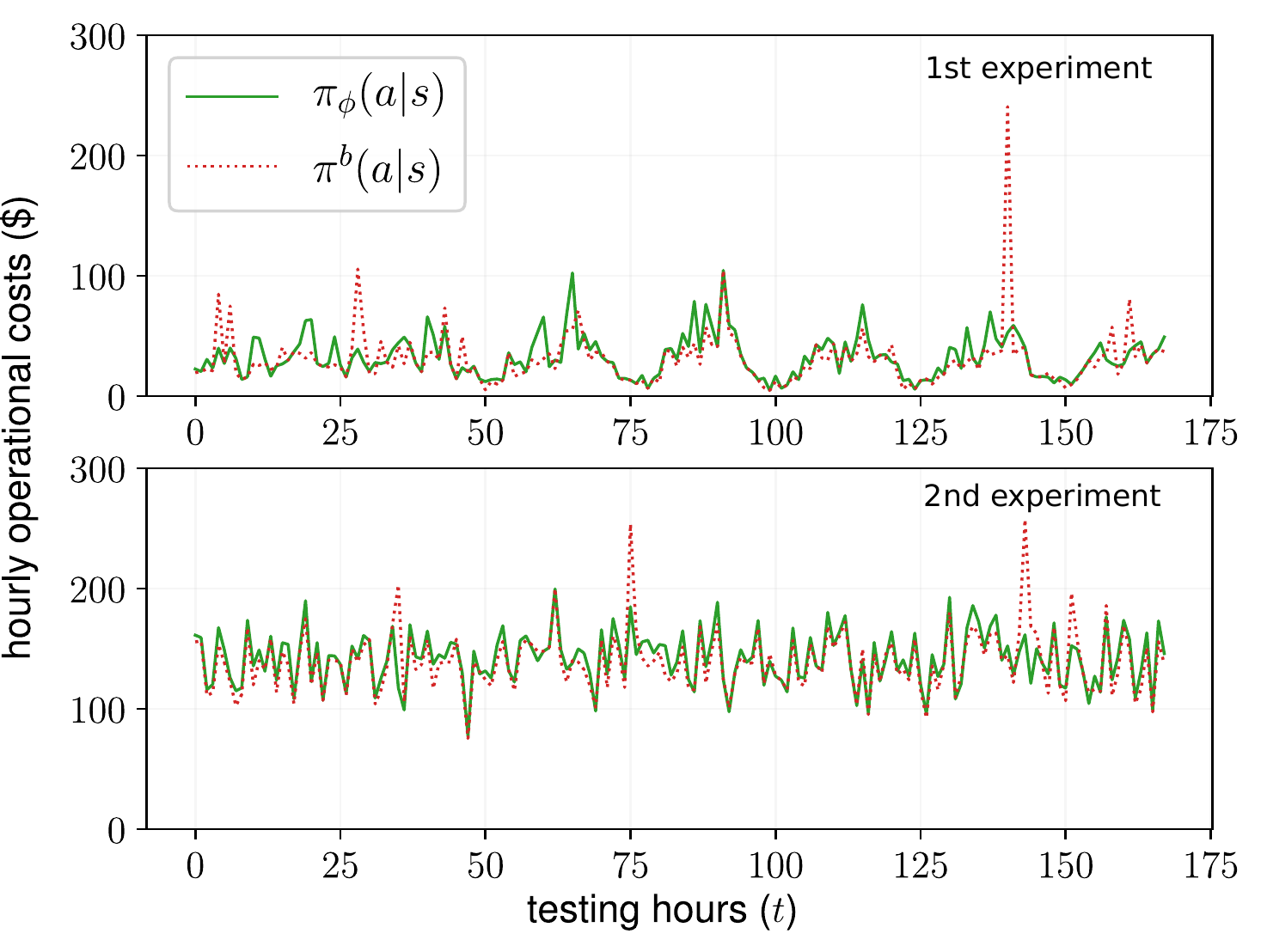}
	\caption{BCSAC agent's response to unforeseen states}
	\label{figtestunusual}
\end{figure}

The dynamic DNR results for the first and second experiments are shown in upper and lower half of Fig. \ref{figtestunusual}, respectively. 

The green curve represents results from the BCSAC agent; the dotted red curve is the behavior policy defined by $[P_{mod}, P_{fix}, P_{rnd}]=[0.7, 0.24, 0.06]$. Fig. \ref{figtestunusual} shows that even if the network configuration is new or the power injection pattern is unfamiliar, the trained BCSAC agent yields lower or nearly the same operational cost as that of the behavior policy. These two experiments show that the BCSAC algorithm is robust against unfamiliar or extreme scenarios.

\subsection{Computation Speed}
This subsection demonstrates the superior computation speed of RL-based control over the model-based control methods. We adopt the MPC-based dynamic DNR algorithm \cite{Dorostkar2016Value} as the model-based benchmark. The MICP is implemented in MATLAB with YALMIP optimization modeling toolbox \cite{Lofberg2004} and MOSEK 9.1 optimization solver. The reinforcement learning algorithms are implemented in Python with TensorFlow 1.14 deep learning framework. They are executed on a desktop with a 4-core Intel i5 3.3GHz CPU and an Nvidia GeForce GTX 1060 GPU. The training and testing time of all methods are provided in Table \ref{tablespeed}.

\begin{table}[h]
	\caption{Total Computation Time of Testing Week}
	\begin{tabular}{p{1cm} p{1cm} R{1.cm} R{1.cm} R{1.cm} R{1.cm}}
		\toprule  
		 & & 16-bus & 33-bus & 70-bus & 119-bus \\ \hline
		\multirow{3}{*}{\specialcell[t]{Training\\(seconds)}} 
		    & DQN & 15.3 & 22.4 & 46.2 & 71.5 \\ 
		    & SAC & 84.2 & 107.4 & 204.3 & 410.9 \\ 
		    & BCSAC & 86.3 & 112.7 & 312.4 & 907.4 \\ 
		    & CVAE & 222.0 & 664.5 & 2040.0 & 6487.1 \\ \hline
		\multirow{3}{*}{\specialcell[t]{Testing\\(seconds)}} 
		    & DQN & 0.2 & 0.4 & 0.9 & 2.1 \\ 
		    & SAC & 0.2 & 0.4 & 1.0 & 2.2 \\ 
		    & BCSAC & 0.2 & 0.4 & 0.9 & 2.2 \\
		    & \multicolumn{2}{l}{MICP MPC} & & & \\
		    & $\;H=1$ & 63.1  & 241.6 & 533.9 & 1341.4 \\
		    & $\;H=2$ & 143.6  & 2439.8 & 8397.2 & $-$ \\
		    & $\;H=5$ & 876.3  & $-$ & $-$ & $-$ \\
		\bottomrule
	\end{tabular}
	\label{tablespeed}
\end{table}
The training of RL-based algorithms can be done in an off-line manner. Therefore, it is more meaningful to compare the testing time of RL-based and model-based control algorithms. As shown in Table \ref{tablespeed}, the computation time of RL-based algorithms are at least two orders of magnitudes shorter than the model-based control algorithms. The advantage of the RL-based algorithms becomes more pronounced when the size of the distribution network increases. With an optimization horizon $H$ of 5 hours, the optimization solver of the model-based controller fails to converge within one hour.

\section{Conclusion} 

This paper proposes a batch-constrained reinforcement learning algorithm to solve the dynamic distribution network reconfiguration problem. Although state-of-the-art off-policy reinforcement learning algorithms have shown great promise as controllers for power distribution systems, they can have lackluster performance when the training dataset is uncorrelated to the true distribution under the current policy or when the state and action domains are extremely large. To learn an effective control policy for dynamic distribution network reconfiguration problems from a limited historical operational dataset, we develop a batch-constrained soft actor-critic (BCSAC) algorithm, which is trained to minimize both the system operational cost and the discrepancy between the policy under evaluation and the historical operational strategy.

Comprehensive test results on four distribution networks show that the proposed BCSAC algorithm not only outperforms state-of-the-art off-policy RL algorithms but also outmatches or achieves similar level of performance as that of the behavior control policy without any information about the network parameters. The proposed algorithm is also very scalable and has much lower computation time than model-based controllers. 

In the future, we plan to develop an asymptotic constraint-satisfaction policy in the batch reinforcement learning setup. We also plan to improve the batch-constrained soft policy iteration theory by exploring different reward regularization mechanisms.

\appendices
\section{Proofs}
\subsection{Proof of Lemma \ref{lemma1}}
\begin{proof}
Defining the augmented reward $r^{\pi}(s,a) = r(s,a) - \tau\mathbb{E}_{s^\prime\sim p} D_{\m{KL}}(\pi(\cdot|s^\prime)||\pi^b(\cdot|s^\prime))$, the operator $\mc{T}^\pi$ can be expressed as:
\begin{equation*}
    \mc{T}^{\pi}q(s,a) = r^{\pi}(s,a) + \gamma \mathbb{E}_{s^\prime\sim P}\mathbb{E}_{a^\prime\sim \pi}[q(s^\prime, a^\prime)] 
\end{equation*}
Or in vector notation $\mc{T}^{\pi}q = r^{\pi} + \gamma P^\pi q$,
where the entry of the vector $P^\pi q$ is given by
\begin{equation*}
    (P^\pi q)(s,a) = \int_{\mc{S}\times\mc{A}} q(s^\prime,a^\prime) d P^\pi(s^\prime,a^\prime|s,a)
\end{equation*}
$P^\pi(s^\prime,a^\prime|s,a) = p(s^\prime|s,a)\pi(a^\prime|s^\prime)$. As $D_{\m{KL}}(\pi(\cdot|s)||\pi^b(\cdot|s))$ is assumed to be bounded for all $s$, $r^{\pi}(s,a)$ is bounded for all $s,a$. Therefore, for any $q(s,a)\in \mathbb{R}, q^\prime(s,a)\in \mathbb{R}, \forall (s,a)\in \mc{S}\times \mc{A}$:
\begin{align*}
    ||\mc{T}^{\pi}q - \mc{T}^{\pi}q^\prime||_\infty
    = \gamma ||P^\pi(q - q^\prime)||_\infty
    \leq \gamma ||q - q^\prime||_\infty
\end{align*}
The inequality is due to $||P^\pi||_\infty = 1$. Therefore, for any $\gamma<1$, $\mc{T}^\pi$ is a contraction mapping with respect to the supremum norm. By the Banach fixed point theorem, the operator $\mc{T}^\pi$ has a unique fixed point $q_\pi^{\m{d}}$ and the sequence defined by $q^{k+1} = \mc{T}^\pi q^k$ converges to this fixed point as $k\rightarrow \infty$.
\end{proof}

\subsection{Proof of Lemma \ref{lemma2}}
\begin{proof}
Since $\pi^\prime(\cdot|s)$ is a maximizer of the objective function $J_{\pi}(\tilde{\pi}(\cdot|s)) = \mathbb{E}_{a\sim\tilde{\pi}}[q_\pi^{\m{d}}(s,a)] - \tau D_\m{KL}(\tilde{\pi}(\cdot|s)||\pi^b(\cdot|s))$, therefore we have $J_{\pi}(\pi^\prime(\cdot|s)) \geq J_\pi(\pi(\cdot|s))$. Thus
\begin{align}
    \mathbb{E}_{a\sim\pi^\prime}[q_\pi^{\m{d}}(s, &a)] - \tau D_\m{KL}(\pi^\prime(\cdot|s)||\pi^b(\cdot|s)) \geq \nonumber \\ \mathbb{E}_{a\sim\pi}[& q_\pi^{\m{d}}(s,a)] - \tau D_\m{KL}(\pi(\cdot|s)||\pi^b(\cdot|s)) \triangleq v_\pi^\m{d}(s) \label{ineq26}
\end{align}
Let (\ref{ineq26}) holds for every $s\in\mc{S}$, we can obtain the following chain of inequalities by repeatedly invoking (\ref{ineq26}) and (\ref{eq22}):
\begin{align}
    q_\pi^\m{d}(s,a) = r + \gamma \mathbb{E}_{s^\prime\sim P}&[v_\pi^\m{d}(s^\prime)] \nonumber\\
     \leq r + \gamma \mathbb{E}_{s^\prime\sim P}&[\mathbb{E}_{a^\prime\sim\pi^\prime}[ q_\pi^{\m{d}}(s^\prime,a^\prime)] \nonumber \\ &- \tau D_\m{KL}(\pi^\prime(\cdot|s^\prime)||\pi^b(\cdot|s^\prime))] \nonumber \\
     = r + \gamma \mathbb{E}_{s^\prime\sim P}&[\mathbb{E}_{a^\prime\sim\pi^\prime}[r + \gamma \mathbb{E}_{s^{\prime\prime}\sim P}[v^{\m{d}}_\pi(s^{\prime\prime})]] \nonumber \\ &- \tau D_\m{KL}(\pi^\prime(\cdot|s^\prime)||\pi^b(\cdot|s^\prime))] \nonumber \\
     \leq ... \qquad \quad \;\;\; &\; \nonumber \\
     \leq q^{\m{d}}_{\pi^\prime}(s,a) \;\;\;\; &\quad\label{ineq27}
\end{align}
Continuously expanding the terms, we obtain $q^{\m{d}}_{\pi^\prime}(s,a)$ on the right hand side by its definition.
\end{proof}
\subsection{Proof of Theorem \ref{theorem1}}
\begin{proof}
The sequence $q^{\m{d}}_{\pi^k}(s,a), k=1,2,...$ generated by repeated applications of the policy evaluation and improvement is non-decreasing and is bounded above. Thus convergence follows from the monotone convergence principle. Denote $q^{\m{d}}_{\pi^\infty}(s,a)$ as the converged value function and $\pi^\infty$ the associated policy. We need to show that $\pi^\infty$ is indeed optimal. Since at convergence, the policy is no longer changing. Therefore $\pi^\infty(\cdot|s)$ is a maximizer of the objective function $J_{\pi^\infty}(\tilde{\pi}(\cdot|s)) = \mathbb{E}_{a\sim\tilde{\pi}}[q_{\pi^\infty}^{\m{d}}(s,a)] - \tau D_\m{KL}(\tilde{\pi}(\cdot|s)||\pi^b(\cdot|s))$. In other words, $J_{\pi^\infty}(\pi(\cdot|s))\leq J_{\pi^\infty}(\pi^\infty(\cdot|s))$ for any policy $\pi$. By the same token as the proof of Lemma 2, this means that $q_{\pi}^\m{d}(s,a)\leq q_{\pi^\infty}^\m{d}(s,a)$. Since $\pi$ is an arbitrary policy, $\pi^\infty$ is indeed the optimal policy.
\end{proof}

\bibliographystyle{IEEEtran}
\bibliography{BCRL_DNR_arXiv}

\end{document}